\documentclass[aps,prx,twocolumn,superscriptaddress,showpacs,longbibliography]{revtex4-1}

\usepackage{amsmath,amssymb, amsfonts, graphics,epsfig,epstopdf,color,verbatim,ulem,braket,tabularx}
\usepackage[colorlinks,linkcolor=blue,citecolor=blue,urlcolor=blue]{hyperref}
\usepackage{wasysym}

\usepackage{tikz}
\usepackage{pgfplots}
\usepackage{bm}% bold math
\usepackage{setspace}   % controllabel line spacing
\newcommand{\beq}{\begin{equation}}
\newcommand{\eeq}{\end{equation}}
\newcommand{\ba}{\begin{array}{ccc}}
	\newcommand{\ea}{\end{array}}

\def\bea{\begin{eqnarray}}
\def\eea{\end{eqnarray}}

\def\<{\langle}
\def\>{\rangle}

\usepackage{slashed}

\newcommand\bowtieA[1][0.8]{
	\begin{tikzpicture}[scale=#1]
	\draw (-0.5,0.866)--(0.5,-0.866)--(-0.5,-0.866)--(0.5,0.866)--(-0.5,0.866);
	
	\node at (-0.5,0.866) [draw,shape=circle,minimum size=2.7mm,inner sep=0pt,outer sep=0pt,fill=white] {};
	\node at (-0.50,0.866) [inner sep=0,font=\small] {$1$};
	\node at (0.5,0.866) [draw,shape=circle,minimum size=2.7mm,inner sep=0pt,outer sep=0pt,fill=white] {};
	\node at (0.50,0.866) [inner sep=0,font=\small] {$0$};
	\node at (-0.5,-0.866) [draw,shape=circle,minimum size=2.7mm,inner sep=0pt,outer sep=0pt,fill=white] {};
	\node at (-0.50,-0.866) [inner sep=0,font=\small] {$0$};
\node at (0.5,-0.866) [draw,shape=circle,minimum size=2.7mm,inner sep=0pt,outer sep=0pt,fill=white] {};
\node at (0.50,-0.866) [inner sep=0,font=\small] {$1$};
	\end{tikzpicture}
}
\newcommand\pgelA[1][0.8]{
	\begin{tikzpicture}[scale=#1]
	\draw (-0.5,0.866)--(0.5,-0.866)--(-0.5,-0.866)--(0.5,0.866)--(-0.5,0.866);
	\draw [opacity=0.3] (-1,-1.154)--(0,-0.577)--(0,0.577)--(-1,1.154);
	\draw [opacity=0.3] (0,0.577)--(1,1.154);
	\draw [opacity=0.3] (0,-0.577)--(1,-1.154);
	
	\node at (-0.5,0.866) [draw,shape=circle,minimum size=2.7mm,inner sep=0pt,outer sep=0pt,fill=white] {};
	\node at (-0.50,0.866) [inner sep=0,font=\small] {$1$};
	\node at (0.5,0.866) [draw,shape=circle,minimum size=2.7mm,inner sep=0pt,outer sep=0pt,fill=white] {};
	\node at (0.50,0.866) [inner sep=0,font=\small] {$0$};
	\node at (-0.5,-0.866) [draw,shape=circle,minimum size=2.7mm,inner sep=0pt,outer sep=0pt,fill=white] {};
	\node at (-0.50,-0.866) [inner sep=0,font=\small] {$0$};
	\node at (0.5,-0.866) [draw,shape=circle,minimum size=2.7mm,inner sep=0pt,outer sep=0pt,fill=white] {};
	\node at (0.50,-0.866) [inner sep=0,font=\small] {$1$};
	\node at (0,0.577) [inner sep=0] {$\scriptstyle \sigma_I^z$};
	\node at (0,-0.577) [inner sep=0] {$\scriptstyle \sigma_J^z$};
	\end{tikzpicture}
}
\newcommand\pgelB[1][0.8]{
	\begin{tikzpicture}[scale=#1]
	\draw (-0.5,0.866)--(0.5,-0.866)--(-0.5,-0.866)--(0.5,0.866)--(-0.5,0.866);
	\draw [opacity=0.3] (-1,-1.154)--(0,-0.577)--(0,0.577)--(-1,1.154);
	\draw [opacity=0.3] (0,0.577)--(1,1.154);
	\draw [opacity=0.3] (0,-0.577)--(1,-1.154);
\node at (-0.5,0.866) [draw,shape=circle,minimum size=2.7mm,inner sep=0pt,outer sep=0pt,fill=white] {};
\node at (-0.50,0.866) [inner sep=0,font=\small] {$0$};
\node at (0.5,0.866) [draw,shape=circle,minimum size=2.7mm,inner sep=0pt,outer sep=0pt,fill=white] {};
\node at (0.50,0.866) [inner sep=0,font=\small] {$1$};
\node at (-0.5,-0.866) [draw,shape=circle,minimum size=2.7mm,inner sep=0pt,outer sep=0pt,fill=white] {};
\node at (-0.50,-0.866) [inner sep=0,font=\small] {$1$};
\node at (0.5,-0.866) [draw,shape=circle,minimum size=2.7mm,inner sep=0pt,outer sep=0pt,fill=white] {};
\node at (0.50,-0.866) [inner sep=0,font=\small] {$0$};
	\node at (0,0.577) [inner sep=0] {$\scriptstyle -\sigma_I^z$};
	\node at (0,-0.577) [inner sep=0] {$\scriptstyle -\sigma_J^z$};
	\end{tikzpicture}
}
\newcommand\bowtieB[1][0.8]{
	\begin{tikzpicture}[scale=#1]
	\draw (-0.5,0.866)--(0.5,-0.866)--(-0.5,-0.866)--(0.5,0.866)--(-0.5,0.866);
	
	\node at (-0.5,0.866) [draw,shape=circle,minimum size=2.7mm,inner sep=0pt,outer sep=0pt,fill=white] {};
	\node at (-0.50,0.866) [inner sep=0,font=\small] {$0$};
	\node at (0.5,0.866) [draw,shape=circle,minimum size=2.7mm,inner sep=0pt,outer sep=0pt,fill=white] {};
	\node at (0.50,0.866) [inner sep=0,font=\small] {$1$};
	\node at (-0.5,-0.866) [draw,shape=circle,minimum size=2.7mm,inner sep=0pt,outer sep=0pt,fill=white] {};
	\node at (-0.50,-0.866) [inner sep=0,font=\small] {$1$};
	\node at (0.5,-0.866) [draw,shape=circle,minimum size=2.7mm,inner sep=0pt,outer sep=0pt,fill=white] {};
	\node at (0.50,-0.866) [inner sep=0,font=\small] {$0$};
	\end{tikzpicture}
}
\begin{document}

\title{Bosonic SET to SPT Transition under Dyonic LSM Theorem}
\author{Yan-Cheng Wang}
\affiliation{School of Physical Science and Technology, China University of Mining and Technology, Xuzhou 221116, China}
\affiliation{Beijing National Laboratory for Condensed Matter Physics, and Institute of Physics, Chinese Academy of Sciences, Beijing 100190, China}
\author{Xu Yang}
\affiliation{Department of Physics, Boston College, Chestnut Hill, MA 02467, USA}
\author{Ying Ran}
\affiliation{Department of Physics, Boston College, Chestnut Hill, MA 02467, USA}
\author{Zi Yang Meng}
\affiliation{Beijing National Laboratory for Condensed Matter Physics, and Institute of Physics, Chinese Academy of Sciences, Beijing 100190, China}
\affiliation{CAS Center of Excellence in Topological Quantum Computation and School of Physical Sciences,
University of Chinese Academy of Sciences, Beijing 100190, China}
\affiliation{Songshan Lake Materials Laboratory, Dongguan, Guangdong 523808, China}

\begin{abstract}
Employing large-scale quantum Monte Carlo simulatoins, we study the phase diagram of a quantum spin model which is subject to the recently developed dyonic Lieb-Shultz-Mattis (LSM) theorem. The theorem predicts there are symmetry enriched/protected topological (SET/SPT) phases in the phase diagram. Our numerical results reveal a first order quantum phase transition between SET and SPT phases, consistent with an anyon condensation mechanism that enforces SPT phase according to the theorem. Also there exists in the phase diagram a symmetry-breaking phase in the form of superfluid (SF). The transition between SET and SF is continuous and that between SPT and SF is first order. Interestingly, the SET, SPT and SF phases meet at a critical endpoint, whose presence can be universally explained via theory contains emergent gauge field coupled to vortex fields, and consequently reveals the exotic feature of our model even beyond the realm of dyonic LSM.
\end{abstract}

\date{\today} 
\maketitle

\section{Introduction}
\label{sec:intro}

The Lieb-Schultz-Mattis (LSM) theorem~\cite{LSM} and its generalizations to higher dimensions~\cite{OshikawaLSM,HastingsLSM} state that a spin system with fractional spin per unit-cell cannot have a symmetric short-range-entangled ground state, \textit{i.e.}, a non-degenerate gapped ground state on a torus preserving  all the symmetries of the system in the thermodynamic limit. This indicates that a gapped ground state must be something unusual: it can be either a spontaneous-symmetry breaking phase or a symmetry-enriched topological (SET) phase\cite{wen2002quantum,mesaros2013classification,barkeshli2014symmetry} (for example, gapped quantum spin liquids~\cite{lee2006doping,balents2010spin,savary2016quantum,zhou2017quantum} ), the latter of which features fractional excitations and is of much theoretical and experimental interests~\cite{Han2012,YSLeeNMR_aps,Feng2017,Wen2017,Wei2017,Feng2018a,Feng2018b,GYSun2018,Becker2018}. Therefore, LSM theorem is not only of theoretical importance but also is a powerful guiding principle for the search of SET phases in materials.

Recently, another type of novel state of matter, which also draws much attention, is the symmetry-protected topological (SPT) phases, which are symmetric short-range-entangled states but feature anomalous edge states (either gapless or symmetry-breaking)\cite{kane2005z,kane2005quantum,bernevig2006quantum,konig2007quantum,fu2007topological,chen2009experimental,xia2009observation,chen2010local,chen2013symmetry}. Examples of SPT phases include the Haldane phases in integer spin chains~\cite{Haldane1983} and topological insulators~\cite{Kane2010,qi2011topological}. By far most of the experimental realizable SPT phases are essentially captured by free-fermion band theories. Bosonic SPT phases, on the other hand, require intrinsically strong interactions and are much harder to be found experimentally in spatial dimension higher than one\cite{senthil2013integer,liu2014microscopic}. Despite the proposals of realizing bosonic SPT in interacting fermionic systems~\cite{Slagle2015,Bonafide2016,Visualizing2016,BilayerGraphene2017}, a generic guiding principle for SPT phases, such as the LSM for SET phases, is still missing.

\begin{figure}[htp!]
\centering
\includegraphics[width=\columnwidth]{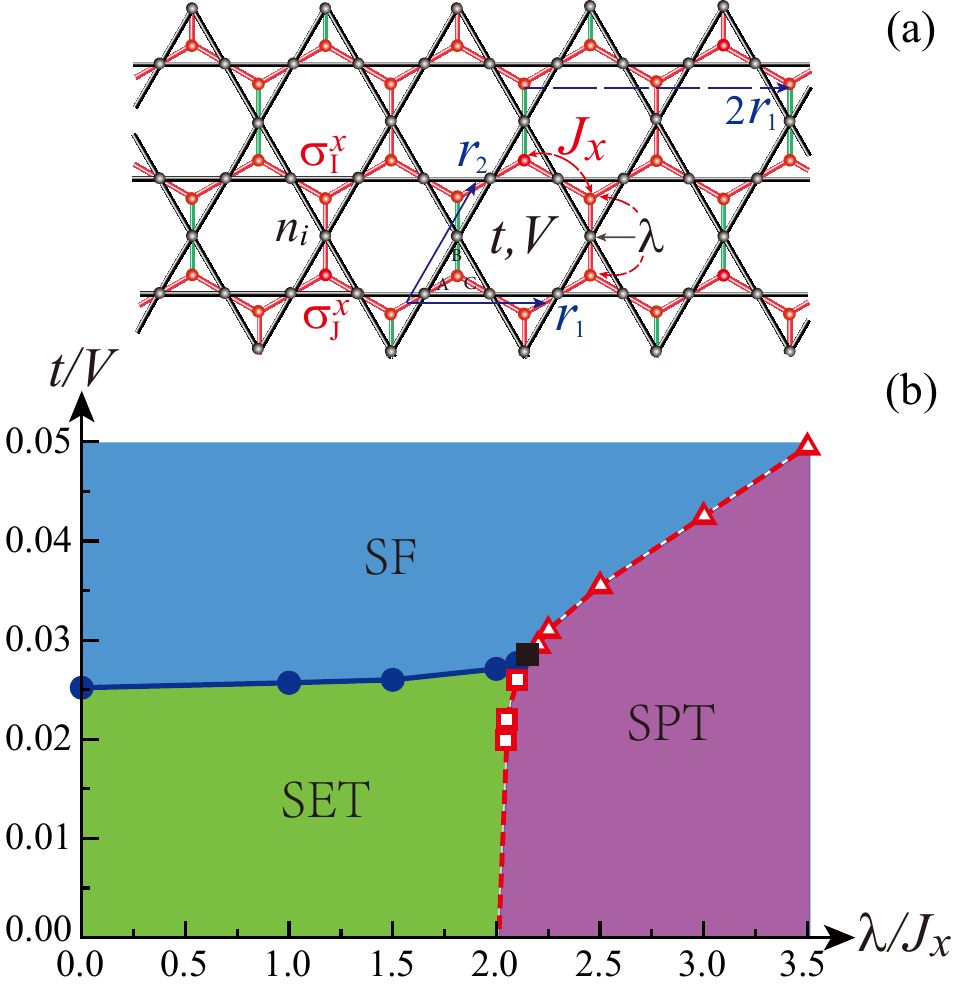}
\caption{(a) The model in Eq.~\eqref{eq:decoratedBFG_full}. The $H^{\text{BFG}}$ is the black kagome layer with boson interactions $t$ and $V$. The $H^{\text{Ising}}$ is the honeycomb layer with Ising interaction $J_x$, and $H^{\text{binding}}$ binds the two layers with $\lambda$ interaction. The green (red) bonds stand for $s_{IJ}=-1(1)$. The $\lambda$ interaction is frustrated in the sense that $\prod\limits_{I,J\in\hexagon}s_{IJ}=-1$. $\mathbf{r}_{1,2}$ are the translational vectors for the kagome lattice unit cell $\mathbf{r}_{1}\times \mathbf{r}_{2}$. The magnetic unit cell for the system in Eq.~\eqref{eq:decoratedBFG_full} is spanned by $2\mathbf{r}_{1}\times \mathbf{r}_{2}$. (b) Ground state phase diagram of model in Eq.~\eqref{eq:decoratedBFG_full}, spanned by the axes $t/V$ and $\lambda/J_{x}$. We set $V=1$, $J_x=0.05$ and tune $t$ and $\lambda$. The transitions between SET and SPT and between SPT and SF are first order, and the transition between SET and SF is continuous. The critical endpoint, where the three phases meet (and consequently is also a triple point), is labeled with solid black dot.}
\label{fig:fig1}
\end{figure}

This is the question we address in this paper. We provide physical guidelines similar to LSM theorem for realizing bosonic SPT phases. Although the original LSM theorems tells nothing about symmetric short-range-entangled phases, recent studies~\cite{MCheng2016,XuYang2017,lu2017lieb} provide generalized LSM theorems, making substantial progress in this direction. The dyonic LSM theorem in Ref.~\onlinecite{XuYang2017} shows that for a system with magnetic translation symmetry and fractional spin in original unit-cell but integer spin in the magnetic unit-cell, besides the usual fates such as symmetry-breaking phases or SET phases, a symmetric short-range-entangled ground state is also possible when certain group algebraic condition is satisfied. Interestingly, when this happens, the ground state is necessarily a nontrivial SPT phase (a phenomenon dubbed symmetry-enforced SPT phases). In this paper, we design a simple model that embodies the requirement of the dyonic LSM theorem and employ large-scale quantum Monte Carlo simulation to solve it and verify the prediction of the theorem. In this way, we provide a concrete example of the existence of bosonic SPT phase in simple and realistic model.

A heuristic understanding of the SPT phases enforced by the dyonic LSM theorem is to start from an SET phase via an anyon condensation mechanism~\cite{jiang2017anyon,lu2017lieb}. In the absence of symmetries in 2+1 dimensions, it is known that condensing (self-statistics) bosonic anyons could confine the topological order. In the context of discrete Abelian gauge theories this is the well studied deconfinement-confinement phase transition~\cite{fradkin1979phase}. New results in Refs.~\cite{jiang2017anyon,duivenvoorden2017entanglement} are, in the presence of symmetries, condensing anyons carrying symmetry quantum numbers would necessarily drive certain SET phases into symmetric short-range-entangled phases, and the dyonic LSM theorem enforces anyons to carry non-trivial symmetry quantum numbers, rendering the obtained symmetric short-range-entangled state to be non-trivial SPT phase.

Under these considerations, we explore the phase diagram of a simple model satisfying the condition of the dyonic LSM theorem. We start from a simple  setting realizing SET phases: the Balents-Girvin-Fisher (BFG) model of half-filled hard-core bosons on the kagome lattice~\cite{BFG2002,sheng2005numerical,Isakov2006,Isakov2007, YCWang2017a,YCWang2017b,GYSun2018}. Based on the BFG model, the dyonic LSM theorem~\cite{XuYang2017} suggests a rather simple microscopic model to realize a bosonic SPT phase via the anyon condensation mechanism. Exactly how the anyon condensation transition happens is beyond the predicting power of the theorem, one has to rely on unbiased numerical simulation, and that is the achievement of this work. 

We investigate the full quantum (i.e. zero temperature) phase diagram of the model in Eq.(\ref{eq:decoratedBFG_full}) via quantum Monte Carlo (QMC) simulations. The obtianed phase diagram Fig.\ref{fig:fig1}(b) includes a superfluid (SF) phase that breaks $U(1)$ charge symmetry, an SET phase that is the well-studied $Z_2$ quantum spin liquid in the BFG model, and a bosonic SPT phase that is obtained via anyon condensation. All three phases are consistent with the dyonic LSM theorem. While the SF-SET phase transition is second-order, both the SET-SPT and the SF-SPT transitions are found to be first-order within the explored parameter regime. And the SF-SET second-order transition is terminated at a critical endpoint on the first order line. Interestingly, we show that the nature of these phase transitions can be understood within a simple mean-field theory, which predicts that the presence of a first-order line and a critical endpoint is universal in the vincinity of the SF-SET second-order phase transition.

More importantly, we found the SET-SPT phase transition acquires three interesting measurable consequences: i) the vison condensation phenomenon that drives the SET-SPT transition is manifested in the bulk vison-vison correlation function; ii) a robust bulk topological index that measures the charge carried by the symmetry defect can be readily computed in the QMC simulation, this topological index takes different values deep in SET and SPT phases and has a jump at phase transition point; iii) the SET-SPT transition has unusual behaviour on the edge: while the SET phase has gapped symmetric edge states, the condensation of vison induces an Ising symmetry breaking transition on the edge of the SPT phase. All these features are found in our QMC simulations, and the main content of the paper is to reveal them in a step by step manner.

\section{Model and dyonic LSM}

Inspired by Ref.~\cite{XuYang2017}, our Hamiltonian is designed by decorating the BFG model with an Ising layer and introducing the coupling between them, as shown in Fig.~\ref{fig:fig1} (a),

\begin{equation}
\label{eq:decoratedBFG_full}
H=H^{\text{BFG}}+H^{\text{Ising}}+H^{\text{Binding}}.
\end{equation}

The BFG model is written in the hard-core boson langurage, 
\begin{equation}
H^{\text{BFG}}=-t\sum_{(i,j)}\left(b_i^\dag b_j + h.c.\right)
+V\sum_{(i,j)}\left(n_i-\frac{1}{2}\right)\left(n_j-\frac{1}{2}\right)
\label{eq:model_BFG}
\end{equation}
slightly different from the original BFG model~\cite{BFG2002}, here the summation $(i,j)$ stands for nearest-, next- and third-neighbor hoppings and interactions. Previous QMC simulations reveal a $Z_2$ quantum spin liquid ground state of $H^{\text{BFG}}$ when $t/V \le 0.025$~\cite{Isakov2006}.

The $H^{\text{Ising}}$ is given by
\begin{equation}
H^{\text{Ising}}=-J_x\sum\limits_{\langle I,J\rangle}\sigma_I^x\sigma_J^x,
\end{equation}
where the Ising degree of freedom lives on the center of every triangle of the Kagome lattice, which comprises a honeycomb lattice and $J_x$ is ferromagnetic interaction between nearest Ising spins.

The binding Hamiltonian is
\begin{equation}
\label{eq:ham_binding}
H^{\text{Binding}}=-\lambda\sum_{
\begin{tikzpicture}
\draw [-][very thick] (0,0) -- (0.5,0); 
\draw [fill] (0,0) circle [radius=.05];
\node at (0,-0.3) {I};
\draw [fill] (0.5,0) circle [radius=.05]; 
\node at (0.5,-0.3) {J};
\node at (0.25,0.25) {i};
\end{tikzpicture}
}
\left(n_i-\frac{1}{2}\right)\left(s_{IJ}\sigma^z_I\sigma^z_J\right),
\end{equation}
where the summation is over all the bonds $\langle I,J \rangle$ on honeycomb lattice with the boson density $(n_{i}-\frac{1}{2})$ at the bond center. The sign $s_{IJ}=\pm 1$ are frustrated in the sense that $\prod\limits_{I,J\in\hexagon}s_{IJ}=-1$. In Fig.~\ref{fig:fig1} (a), a specific choice of $s_{IJ}$ is shown such that red bonds have $s_{IJ}=1$ and green bonds has $s_{IJ}=-1$. The binding term therefore binds  Ising happy bonds ($s_{IJ}\sigma_I^z\sigma_J^z=+1$) and Ising unhappy bonds ($s_{IJ}\sigma_I^z\sigma_J^z=-1$) on a hexegon of the honeycomb layer when $\lambda$ is sufficiently large, and this is the constraint the frustrated $s_{IJ}$ ensures.

It is convenient to introduce the equivalent spin-1/2 description for the hard-core boson degrees of freedom on the kagome lattice: $S_i^z\equiv(n_i-1/2)$.  The important on-site symmetries of the model in Eq.(\ref{eq:decoratedBFG_full}) include the global Ising symmetry $Z_{2I}$ generated by $\prod_I \sigma_I^x$, and an $O(2)=U(1)_b\rtimes Z_{2C}$ symmetry group, where the $U(1)_b$ boson number conservation symmetry is generated by $\prod_i e^{i\theta S_i^z}$ and $Z_{2C}$ is generated by $\prod_i S_i^x\prod_{I\in A}\sigma_I^x$ (the last product is over sites in the A-sublattice on the honeycomb lattice). This $O(2)$ symmetry sharply defines a projective representation (i.e. a half-integer spin) per original unit cell in model Eq.(\ref{eq:decoratedBFG_full}). Besides the on-site symmetry group, the system also has spatial symmetries. Due to the frustrated nature of the binding term, the original translation operations $T_x^{orig.}$ and $T_y^{orig.}$ of the kagome layer of $H^{\text{BFG}}$ shall now be combined with certain Ising symmetries to be the magnetic translations $T_x$ and $T_y$ of the entire system, which satisfies the following magnetic translation algebra:

\begin{equation}
T_xT_yT_x^{-1}T_y^{-1}=\prod\limits_{I} \sigma_I^x.
\label{eq:mag_trans}
\end{equation}

A simple counting shows that each original unit-cell hosts 3/2 bosons, corresponding to a half-integer $O(2)$-spin, and the magnetic unit-cell has 3 bosons (the original unit-cell is spanned by $(\mathbf{r}_{1}, \mathbf{r}_{2})$ and the magnetic unit-cell is spanned by $(2\mathbf{r}_{1}, \mathbf{r}_{2})$, as shown in the Fig.~\ref{fig:fig1} (a)), such counting satisfies the requirement of the dyonic LSM theorem. Therefore, based on the generic conclusion in Ref.~\cite{XuYang2017}, if our model in Eq.(\ref{eq:decoratedBFG_full}) leads to a symmetric short-range-entangled state, it must be a nontrivial SPT state protected by the $Z_{2I}\times O(2)$ symmetry. In fact, $Z_{2I}\times U(1)_b$ symmetry is enough to protect such a nontrivial SPT state.

A heuristic picture of how dyonic LSM works in our case is as follows. 
One way of obtaining short-range entangled state is to start from the SET phase and condense bosonic anyons. In this way, we can confine the topological order completely, and if we require symmetries to be unbroken, those condensed anyons should carry trivial symmetry fractionalizations. For the $Z_2$ spin liquid phase in the original BFG model, such a short-range-entangled state is impossible since both the spinons and visons carry symmetry fractionalizations. We cannot condense spinon without breaking the $U(1)_b$ symmetry since it has fractional $U(1)_b$ quantum number. And the visons also has symmetry fractionalization under translation: when translated around a unit-cell, a vison will pick up a $-1$ Berry phase resulting from the braiding with the background spinon residing in each unit-cell~\cite{GYSun2018}. Therefore in this case condensing visons will necessarily break lattice translation symmetry and result in valence bond ordered phase, as explicitly shown in Ref.~\onlinecite{YCWang2017b,GYSun2018}.

However, in the full model in Eq.~\eqref{eq:decoratedBFG_full}, the magnetic translation (Eq.~\eqref{eq:mag_trans}) is introduced, which allows vison to carry trivial symmetry fractionalization. Under magnetic translation, vison is not only translated around a unit-cell, but also acted upon by the global Ising symmetry. If vison is odd under Ising symmetry, it will get an extra $-1$ phase under Ising symmetry to cancel the $-1$ Berry phase resulting from the braiding with background spinons, rendering the vison symmetry fractionalization under magnetic translation to be trivial. Therefore, condensing such Ising-odd visons will preserve both the on-site and translation symmetries. 

According to the anyon condensation mechanism in Ref.~\onlinecite{jiang2017anyon,XuYang2017}, condensing visons carrying odd Ising quantum numbers will necessarily result in a nontrivial SPT state. To see this, we can couple the $Z_{2I}$ Ising symmetry to a $Z_2$ dynamical gauge field (we call it $Z^g_{2I}$ below) in the vison-condensed phase, which is the familiar gauging procedure. The Ising-odd vison corresponds to the bound state of vison and $Z^g_{2I}$ gauge charge in the gauged state and is condensed. This bound state of $Z^g_{2I}$ gauge flux and spinon, a \emph{dyonic} object, will have trivial mutual statistics with the condensed vison and remains as the un-confined excitations. Upon un-gauging, this dyon becomes the $Z_{2I}$ Ising symmetry defect carrying half-integer $U(1)_b$ charge, which indicates that the state is indeed a non-trivial SPT state. In addition, the dyonic LSM theorem proposed in Ref.~\onlinecite{XuYang2017} dictates that such a nontrivial SPT state is the only possible short-range entangled state in the presence of the magnetic translation symmetry and the on-site symmetries.

Although the dyonic LSM theorem is powerful in terms of revealing the possibility of having nontrivial SPT states, exactly in which regime such bosonic SPT phase exists and what is its relations with the SET phase and other symmetry-breaking phases of the model, will have to be addressed via unbiased numerical calculations, and that is the accomplishment of this work.

\section{Structure of the phase diagram}
Before discussing the QMC results, it is helpful to analyze the ground state of the model in certain physical limits. To simplify the discussion, throughout the paper, we fix the parameters $V=1$, $J_x=0.1$ and tune $t$, $\lambda$, such that the phase diagram is spanned by two axes $t/V$ and $\lambda/J_x$, as shown in Fig.~\ref{fig:fig1} (b).

When the kinetic term dominates, $t$ is large compared to $V$, $J_x$ and $\lambda$, interaction becomes negligible and the system will always enter the superfluid (SF) phase. This is the blue area in Fig.~\ref{fig:fig1} (b).

In the other limit where $t\rightarrow 0$, interactions come into play and topological phases emerge. One can easily see that in the decoupling limit with $\lambda=0$, the ground state is the SET state of the BFG model~\cite{Isakov2006,Isakov2011,Isakov2012,YCWang2017a,YCWang2017b}, this is the green area in Fig.~\ref{fig:fig1} (b). If we turn off $t$ completely, the low energy manifold is spanned by degenerate states with 3 boson per kagome plaquette. Small $t$ term will lift the degeneracy and results in the following effective Hamiltonian 

\begin{align}
\label{eq:originalBFG}
H^{\text{BFG}}_{\text{eff}}=&-J_{\text{ring}}\sum\limits_{\bowtie}(\big|\raisebox{-4mm}\bowtieB\big\rangle \big\langle\raisebox{-4mm}\bowtieA\big|+h.c.),
\end{align}
where boson numbers are indicated in the circle and $J_{\text{ring}}=\frac{4t^2}{V}$. It is well-known that the ground state of $H^{\text{BFG}}_{\text{eff}}$ is an SET state~\cite{BFG2002}.

In this SET phase, the gapped topological excitations include the $Z_2$ gauge charge (spinon) $e$ and the gauge flux (vison) $v$. Only the spinon $e$ carrys nontrivial on-site symmetry fractionalization: half $U(1)$-charge. The visons, on the other hand, can carry Ising quantum numbers~\cite{YCWang2017a,YCWang2017b,GYSun2018}. The vison-pair creation operator can be written as the product of boson density $(2n_i-1)$ over a path $C$ on the kagome lattice
\begin{equation}
v_I^ev_J^e\equiv \prod_{i\in C}\!\!\!\!\!\!\!\!\!\!\longrightarrow(2n_i-1),
\label{eq:barevisoncorr}
\end{equation}
which creates two visons at the end points $I$ and $J$ of the vison string $C$, as shown in the path in Fig.~\ref{fig:fig4} (a). Here the superscripts $e$ in $v_I^e$ indicate that visons are even under Ising symmetry. This correlator is short-ranged (exponential decay) in the SET ($Z_2$ quantum spin liquid) phase, as the visons are gapped excitations. This has been explicited demonstrated via QMC simulation in our previous work~\cite{GYSun2018}.

In the strong $\lambda$-coupling regime, we can treat $V$ and $\lambda$ as the largest energy scale and $t$ and $J_x$ as perturbations. Before adding $t$ and $J_x$ terms, the low energy manifold is spanned by states with 3 boson per kagome plaquette satisfying new local constraints involving hard-core bosons and adjacent Ising spins on the honeycomb lattice: \begin{equation}
\label{constraint}
(2n_i-1)(s_{IJ}\sigma^z_I\sigma^z_J)=1.
\end{equation} 
If we turn on $t$ and $J_x$, as shown in the Appendix~\ref{app:Heff}, perturbative calculations reveal that in the parameter regime where $\lambda,V\gg t,J_x$ and $J_x(8+2V/\lambda)\gg t$, we have a low-energy effective Hamiltonian of the entire model in Eq.~(\ref{eq:decoratedBFG_full}) as
\begin{equation}
H_{\text{eff}}=-J_{\text{ring}}\sum_{\bowtie}(\big|\raisebox{-4mm}\pgelB\big\rangle \big\langle\raisebox{-4mm}\pgelA\big|+h.c.),
\end{equation}
where boson numbers are indicated in the kagome site and the two central Ising spins are also flipped to ensure the constraint in Eq.~\eqref{constraint} and 
\begin{equation}
J_{\text{ring}}=\frac{2t^2J(V+4\lambda)}{(V+2\lambda)^2\lambda}.
\end{equation} 
This Hamiltonian can be solved via a mapping between the low energy Hilbert space of $H_{\text{eff}}$ and that of $H^{\text{BFG}}_{\text{eff}}$. Appendix~\ref{app:solveHeff} explains that $H_{\text{eff}}$ has a unique symmetric gapped ground state on torus, therefore it is not the SET ground state of $H^{\text{BFG}}_{\text{eff}}$ with $Z_2$ topological order (four-fold degeneracy on the torus), but as we will discuss later, it is an SPT state with anomalous edge state. It is the purple area in Fig.~\ref{fig:fig1} (b).

Therefore in the small-$t$ case, we know the ground states of our model in the two limits: an SET phase at small $\lambda$ and an SPT phase with $\lambda$ sufficiently large. In the phase diagram of Fig.~\ref{fig:fig1} (b), our QMC simulations show that there exists no intermediate phase between the SET and the SPT states when we tune $\lambda$. Such a direct phase transition between two symmetric states can be understood within the anyon condensation scenario. The form of the coupling in Eq.~\eqref{eq:ham_binding} can in fact be interpreted as the nearest-neighbor Ising-odd-vison hopping term (see Eq.\eqref{eq:visoncorrelator}). Therefore when $\lambda$ is large enough, we expect the Ising-odd visons to condense and hence the $Z_2$ gauge dynamics are confined without symmetry-breaking. As discussed before, this confined phase obtained via anyon condensation must be a nontrivial SPT phase. In particular, the Ising defect in this SPT phase must carry half $U(1)_b$-charge. 

\section{Measurable consequences for the SET-SPT phase transition}
\label{sec:measurable}
In this section, we introduce some measurable quantities in the QMC simulation across the SET-SPT transition, and the results will be shown in the next section. 

One can directly probe the vison condensation phenomenon via the following Ising-odd-vison (denoted by $v^o$) correlator
\begin{equation}
\label{eq:visoncorrelator}
v_I^ov_J^o\equiv\sigma^z_I\sigma^z_J\prod_{i\in C}\!\!\!\!\!\!\!\!\!\!\longrightarrow(2n_i-1),
\end{equation}
where the superscript $o$ in $v_I^o$ indicates it is odd under Ising symmetry.

This is slightly different from the vison-pair correlation in Eq.~\eqref{eq:barevisoncorr} in that here we attach two local operators $\sigma_I^z$ and $\sigma_J^z$ at the end of the string. This is to ensure that visons carry odd $Z_{2I}$ Ising charge. 

In the SET phase, all visons are gapped, therefore the Ising-odd-vison correlator should be short-ranged. In the SPT phase, however, vison-vison correlator is long-ranged. We can take the ground state of $H^{}_{\text{eff}}$ to illustrate this point. Since the constraint Eq.\eqref{constraint} is satisfied everywhere for the low-energy Hilbert space of $H_{\text{eff}}$, by taking product of terms $(2n_i-1)(s_{IJ}\sigma^z_I\sigma^z_J)$ along $C$, we find that $v_Iv_J$ is just equivalent to products of all the $s_{MN}$ with bonds $MN\in C$ in the restricted Hilbert space and hence will receive a constant expectation value in the ground state of $H_{\text{eff}}$, i.e. the SPT state, no matter how far site $I$ and $J$ are separated, this indicates the Ising-odd-visons are condensed in the SPT phase. As shown in the next section, such prediction is consistent with our QMC observation in Fig.~\ref{fig:fig4} (b).

One can also use a topological index to characterize the SET-SPT phase transition. According to the dyonic LSM theorem, the Ising defect in the SPT phase will carry half $U(1)$ charge in order to screen the fractional charge in a magnetic unit-cell. To measure the charge carried by an Ising defect, one can create 2 Ising defects with separation $l$ much larger than the correlation length $\xi$ and measure the total charge  $N_{\mathcal{D}}$ within a region $\mathcal{D}$ with radius $r$ around one Ising defect, where $\xi \ll r\ll l$. The difference between $N_{\mathcal{D}}$ of the ground state and $N_{\mathcal{D}}$ of the state with an Ising defect reveals the charge carried by the Ising defect. 

To this end, a local measurement of the total charge within a region is needed. A natural definition of the total $U(1)$ charge in a plaquette is $\frac{1}{2}\sum\limits_{i\in \hexagon} n_i$, the prefactor $\frac{1}{2}$ is due to the fact that every site is shared by two plaquettes. So for a region $\mathcal{D}$ on the honeycomb lattice, the total charge $N_{\mathcal{D}}$ can be represented by the following expression
\begin{equation}
\label{eq:localbosonnumber}
N_{\mathcal{D}}=\sum\limits_{\hexagon\in \mathcal{D}}(\frac{1}{2}\sum\limits_{i\in \hexagon}n_i).
\end{equation}
The fractional part in $N_{\mathcal{D}}$ can be readily extracted by taking the exponential: $e^{i 2\pi N_{\mathcal{D}}}$, which lead us to the measurement of the following topological index:
\begin{equation}
\label{eq:bosonnumber}
Z=\langle e^{i2\pi N_{\mathcal{D}}}\rangle.
\end{equation}
This index $Z$ will take value $+1$ when the total charge is an integer and $-1$ when the total charge is a half-odd integer.

In practice, we modify the Hamiltonian by choosing a branch cut line and acting Ising symmetry $\sigma^x$ on only one side of those terms that cross the branch cut, in order to create 2 Ising defects at two end points of the branch cut. This is equivalent to changing the sign of $s_{IJ}$ on binding terms that cross the branch cut. Due to the low energy constraint Eq.~\eqref{constraint}, the net result is to change the boson number ($0\leftrightarrow 1$) on sites crossed by the branch cut.

A closer inspection of Eq.~\eqref{eq:bosonnumber} shows that only the change of boson numbers on the boundary $\partial D$ will contribute to changes in $Z$ since $N_{\mathcal{D}}=\frac{1}{2}(\sum\limits_{i\in \partial D}n_i)+[\text{integer}]$. Because the branch cut line crosses $\partial D$ odd number of times, $Z$ will be changed by $-1$ compared to the ground state when the Ising defect is introduced. As will be discussed in the next session, such a jump of topological index at the SET-SPT transition is observed in Fig.~\ref{fig:fig5} (b).

Precisely speaking the vison string operators Eqs.~\eqref{eq:barevisoncorr} and \eqref{eq:visoncorrelator} and the related local $U(1)_b$ charge operator Eqs.~\eqref{eq:localbosonnumber} and \eqref{eq:bosonnumber} are justified only in the limit of $V/t\rightarrow \infty$. This is the limit in which the contraints of 3-boson per plaquette are fulfilled, and consequently the string operators do not cause excitations along the string (similarly the local $U(1)_b$ operator does not cause excitation along its circumference). For a finite $V/t$ the correct operators would be dressed by fluctuations of boson configurations violating the 3-boson per plaquette constraints, and difficult to write down. In our QMC simulations, similar to previous numerical works on the BFG model~\cite{Isakov2006}, we have projected the wavefunction onto the low-energy subspace where 3-boson per plaquette constraints are exactly satisfied before taking the expectation value of these operators.

The SET-SPT phase transition also manifests itself on the edge. Since both phases are symmetric, they cannot be distinguished by a local order parameter in the bulk. However,  boundaries introduce new physics. It is known that there are 2 different kinds of edges for the SET phases~\cite{Bravyi1998,levin2013}, one is obtained by condensing spinon on the boundary, the other by condensing vison. Since the spinon in the SET state carries fractional $U(1)$ charge, the edge will becomes a Luttinger liquid by condensing them (continuous symmetry cannot be broken in 1D). The visons, on the other hand, do not carry any fractional charge~\cite{QiYang2015,QiYang2016} but can carry Ising quantum number, therefore a gapped symmetric edge can be realized by the condensation of visons. In particular, the Ising-even-vison-condensed boundary can be realized between the SET phase and the vacuum, and the Ising-odd-vison-condensed boundary can be realized between the SET phase and the SPT phase.

If we start from the SET phase with a gapped symmetric boundary, \textit{i.e.}, Ising-even-vison-condensed boundary and then tune $\lambda/J_x$ to drive a bulk phase transition, but ensure that the boundary does not have phase transition preceding the bulk phase transition, we will be able to observe spontaneous Ising symmetry breaking on the edge. The reason is as follows: the bulk phase transition is induced by the condensation of Ising-odd-vison particle. As a consequence, the proximity of the boundary to the bulk causes both the condensation of Ising-odd-vison $v^o$ and Ising-even-vison $v^e$ on the boundary. An $v^o$ and an $v^e$ will fuse into a local Ising-odd operator $\langle v^ev^o \rangle\sim \langle \sigma^z \rangle$ which has a non-zero expectation value, therefore causing the Ising symmetry to be spontaneously broken on the boundary. 

Moreover, since the Ising-even-vison particle is always condensed on the boundary, the Ising-Ising correlator should behave just as the Ising-odd-vison correlator during the phase transition. As an interesting observation, we point out that, \emph{assuming} the bulk phase transition is continuous, this bulk-transition-induced boundary Ising symmetry breaking, although happening on the 1+1D boundary, features the critical exponents of the 3D Ising universality class. For instance, the critical exponent $\beta$ for the $Z_{2I}$ order parameter $\langle\sigma^z\rangle$ on the boundary is expected to be $\beta=\beta_{\mbox{3D Ising}}\approx 0.326$. This is because the continuous bulk vison-condensation transition is well-known to be dual to the 3D Ising universality class\cite{Kogut1979} (the fact that visons carry $Z_{2I}$ charge does not modify this universality class). Namely the deconfined(confined) $Z_2$ gauge theory is dual to the Ising paramagnet (ferromagnet). The $Z_{2I}$-odd vison correlators in the bulk, which is not a local order parameter, is dual to the Ising correlator in the Ising universality class. However, as discussed above, on the boundary this $Z_{2I}$-odd vison correlators becomes the local $Z_{2I}$ order parameter via fusing with the condensed $Z_{2I}$-even visons.

\begin{figure}[htp!]
\centering
\includegraphics[width=\columnwidth]{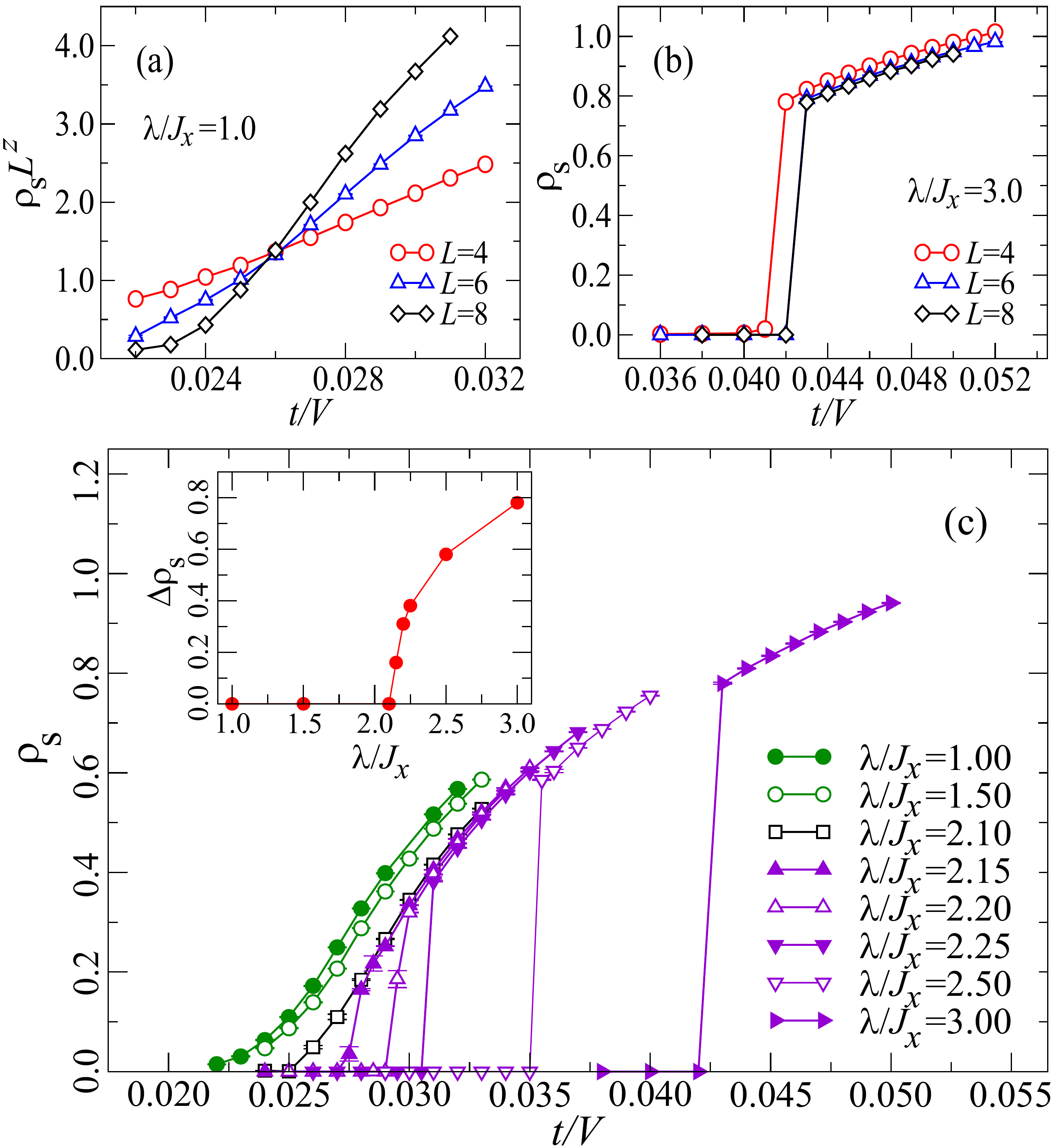}
\caption{(a) The transition from SET to SF as a function of $t/V$ at $\lambda/J_x=1.0$. The superfluid density $\rho_s$ is scaled with the system size $L^{z}$, where $z=1$ is the dynamical exponent of (2+1)D XY$^{*}$ transition~\cite{Isakov2006,Isakov2007,YCWang2017a,YCWang2017b}. The crossing between different system sizes $L=4,6,8$ signifies the transition point. (b) The same analysis for SPT to SF transition at $\lambda/J_x=3.0$, here the SPT to SF transition is obvious first order. (c) $\rho_s$ as a function of $t/V$ for $\lambda/J_x=1$ to $3$ for system size $L=8$. The continuous SET-SF phase transition is terminated around $\lambda/J_x=2.10$, after which the (SPT-SF) transition becomes clearly first order. The inset is the gap (the jump in $\rho_s$) of the superfluid stiffness $\Delta\rho_s$ vs. $\lambda/J_x$.}.
\label{fig:fig2}
\end{figure}

\section{Numerical results}
\label{subsec:numerical results}
Now we are ready to discuss the results obtained from large-scale QMC simulations. To solve the model in Eq.~\eqref{eq:decoratedBFG_full}, we implement a finite-temperature Stochastic Series Expansion (SSE-QMC) alogrithm with directed loop update~\cite{Syljuaasen2002}. Since the model is highly anisotropic and frustrated, i.e., $V \gg t$ in $H^{\text{BFG}}$ and the sign-change of $s_{IJ}$ in $H^{\text{binding}}$, the energy landscape in the configuration space is complicated with many local minima. To overcome the hence induced sampling problem, we perform the QMC update with a 8-spin operator as a plaquette (16 legs in a vortex)~\cite{YCWang2017a,YCWang2017b}, instead of the conventional 2-spin operator. Moreover, to reduce the rejection rate of the proposed spin configuration, we make use of a specific algorithm that satisfies the balance condition without imposing detail balance in the Markov chain of Monte Carlo configurations~\cite{SuwaPRL2010}. Such advanced scheme bestows us the capability of accessing large system sizes and low temperatures. The largest linear system size is $L=16$, note the total lattice site is $N=3\times L \times L \ (\text{kagome})+2\times L \times L \ (\text{honeycomb})$, and we usually set the inverse temperature $\beta=2L/t$ to make sure the finite size systems are at their ground states.

\begin{figure}[htp!]
\centering
\includegraphics[width=\columnwidth]{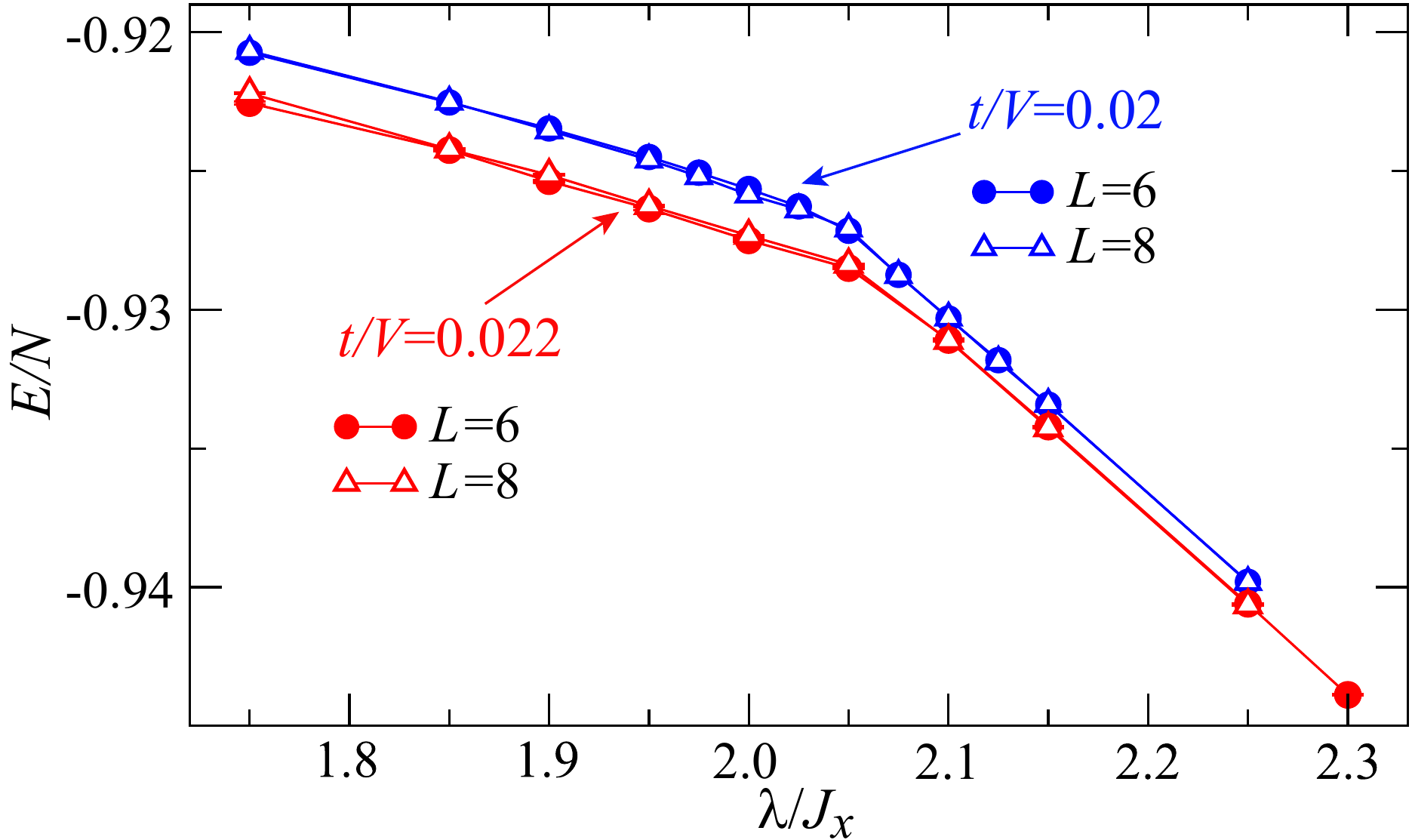}
\caption{(a) The energy density as a function of $\lambda/J_x$ at $t/V=0.022$ and $t/V=0.02$ for $L=6,8$. The cusp signifies the first order transition between SET and SPT phases.}
\label{fig:fig3}
\end{figure}

\begin{figure}[htp!]
\centering
\includegraphics[width=\columnwidth]{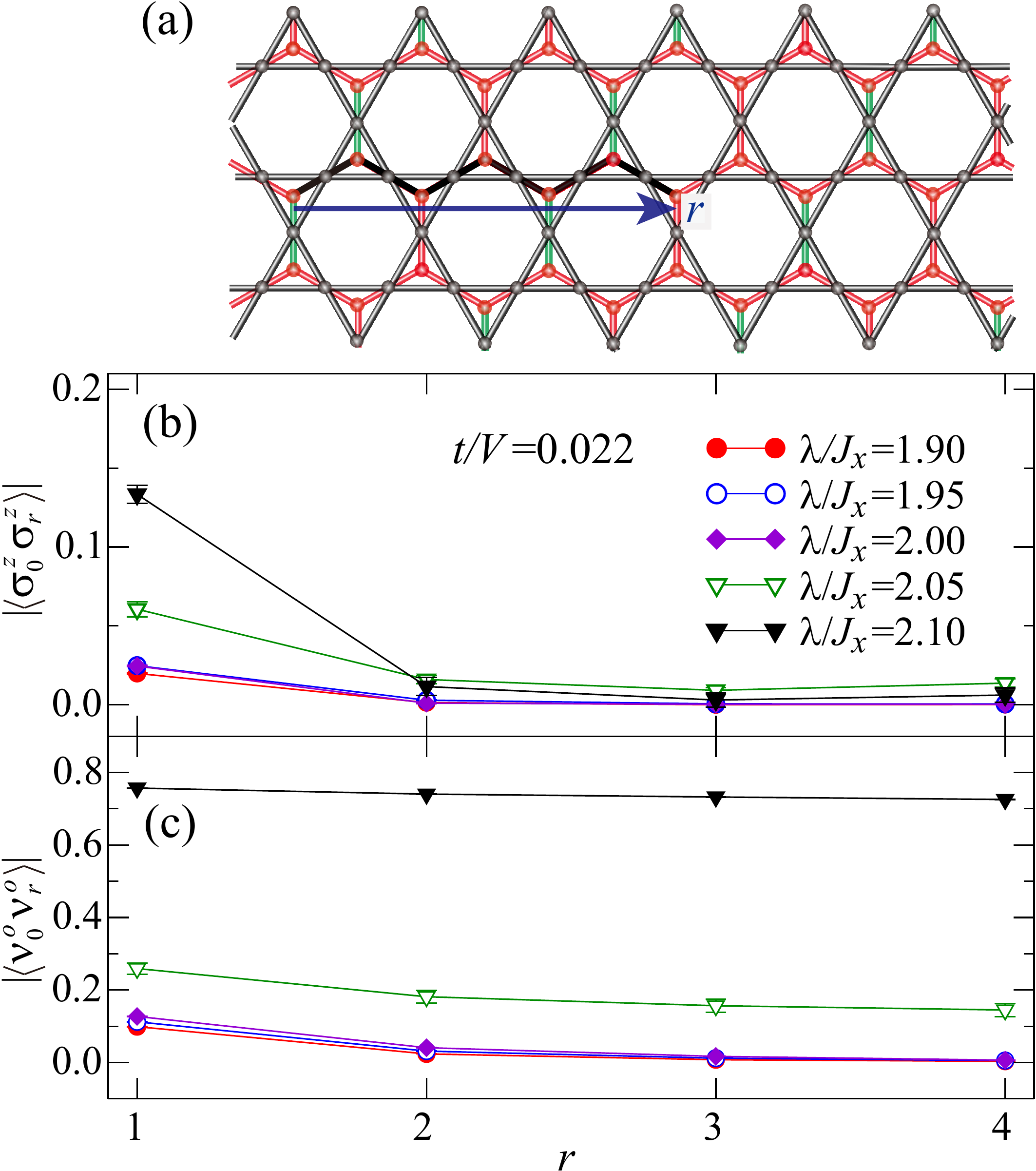}
\caption{(a) Geometry for the calculation of the bulk Ising-odd vison-pair correlation defined in Eq.~\eqref{eq:visoncorrelator}. The black solid line is the path $C$ in Eq.~\eqref{eq:visoncorrelator}. (b) The measured bulk Ising-Ising correlator $|\langle \sigma_0^z \sigma_r^z \rangle|$ as a function of distance $r$. Both in the SET and SPT phase, this correlator is short-ranged. (c) The measured bulk Ising-odd vison-pair correlator $|\langle v_0^o v_r^o \rangle|$ as a function of distance $r$. In the SET phase, when $\lambda/J_x < 2.05$, this correlation is short-ranged and the visons are gapped; in the SPT phase, when $\lambda/J_x \ge 2.05$, this correlation becomes long-ranged, signifying the vison condensation.}
\label{fig:fig4}
\end{figure}

\subsection{Phase diagram}
In QMC simulations, the SF phase is characterized by finite value of superfluid density $\rho_s=\langle W^{2}_{\mathbf{r}_{1}}+W^{2}_{\mathbf{r}_{2}}\rangle/(4\beta t)$ through winding number fluctuations $W^{2}_{\mathbf{r}_{1,2}}$~\cite{Pollock1987}, where $\mathbf{r}_{1,2}$ is the lattice directions, as shown in Fig.~\ref{fig:fig1} (a). $\rho_s$ signifies the onset of off-diagonal long-range order and indicates $U(1)$ symmetry breaking~\cite{Meng2008}. 

\begin{figure}[htp!]
\centering
\includegraphics[width=\columnwidth]{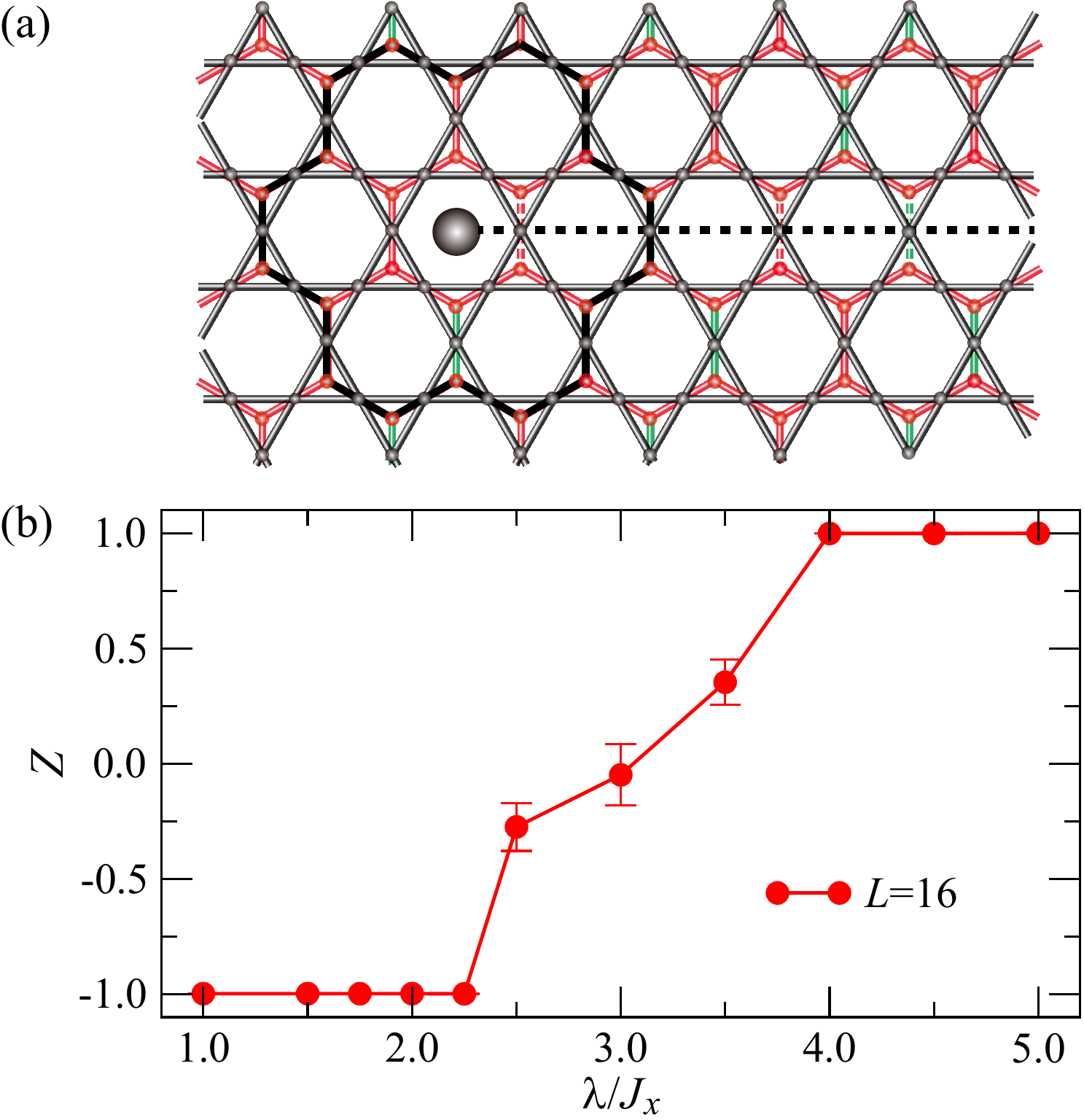}
\caption{(a) Geometry for the calculation of topoological index defined in Eq.~\eqref{eq:bosonnumber}. The black dot stands for the Ising defect and the black solid line circulate the region $\mathcal{D}$ on the honeycomb lattice in which the boson density is counted. (b) The topological index $Z$ as a function of $\lambda/J_x$ at $t/V=0.022$, system size $L=16$. $Z=-1$ in SET phase and $Z=1$ in SPT phase, in between there is a topological phase transition, as shown by the jump of $Z$. The non-integer values of $Z$ are due to finite size effect.}
\label{fig:fig5}
\end{figure}

The phase boundary between SPT and SF phases is determined via the $\rho_s$, the results are shown in Fig.~\ref{fig:fig2}. The SET to SF transition is continuous, as shown in Fig.~\ref{fig:fig2} (a), and this transition is belong to the (2+1)D  XY$^*$ transition with large anomalous dimensition $\eta$ due to the condensation of spinons~\cite{Isakov2011,Isakov2012,YCWang2017b}. The SPT-SF phase transition is first-order, as can be seen from the jump of the superfluid stiffness $\rho_s$ in Fig.~\ref{fig:fig2} (b). 

We found the three phases, SET, SPT and SF, meet at a triple point. The location of this triple point is at $(\lambda/J_x, t/V) \sim (2.1, 0.027)$. This is determined from the $\rho_s$ data in Fig.~\ref{fig:fig2} (c), when the SF-SPT transition is closer to the triple point, the jump of $\rho_s$ will gradually become zero (see Fig.~\ref{fig:fig2} (c)) between $\lambda/J_x=2.15$ and $2.1$, which is a strong indication that the triple point is a critical endpoint. In Sec.\ref{sec:mean-field} we develop a simple mean-field theory to understand the nature of these transitions. It is remarkable that the mean-field theory predicts that the existence of the first-order lines and the critical endpoint is universal.

\begin{figure}[htp!]
\centering
\includegraphics[width=\columnwidth]{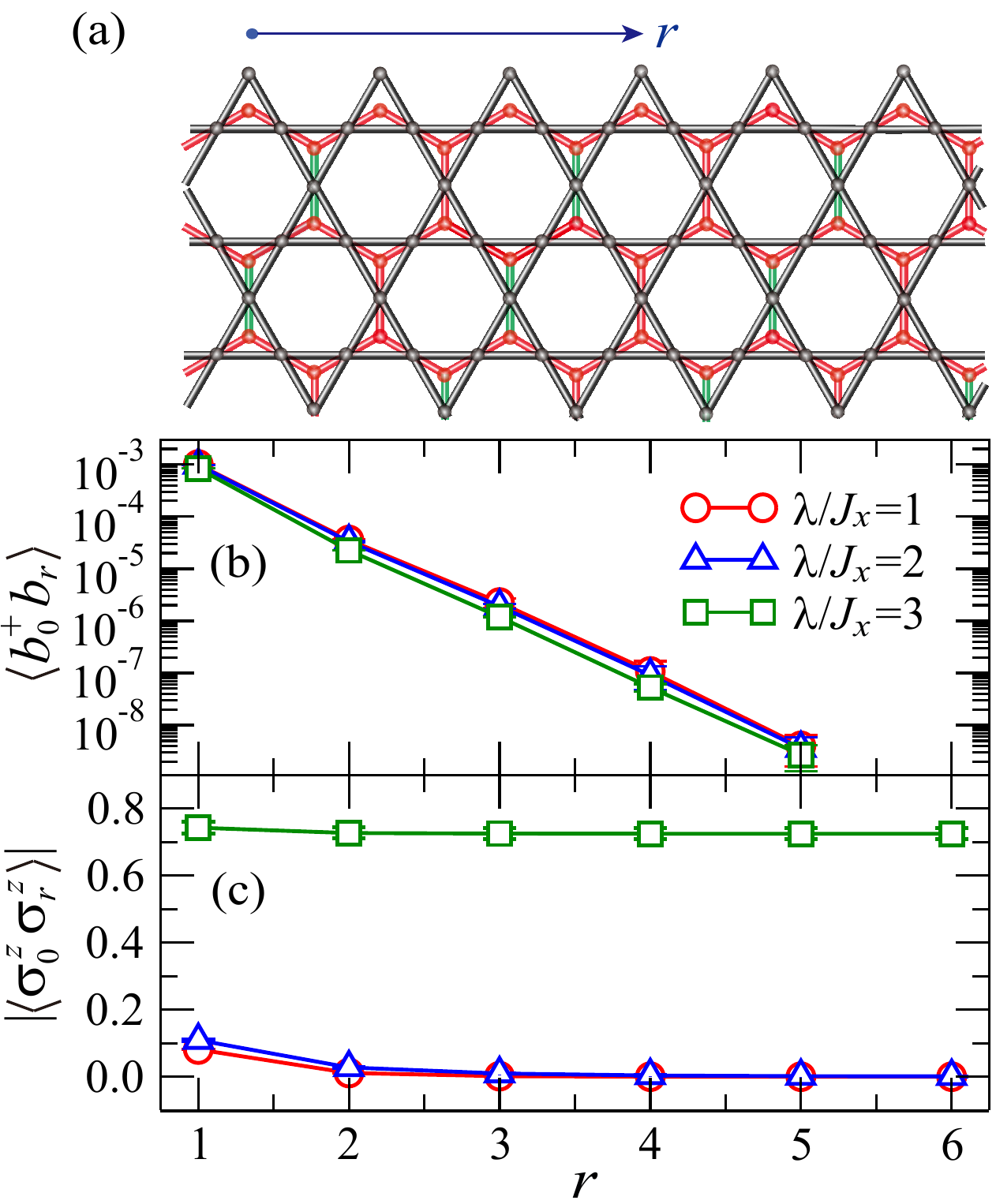}
\caption{(a) The edge of the ribbon geometry for the simulation. (b) The spinon-pair correlation $\langle b^{\dagger}_{r}b_{0} \rangle$ along the edge of the ribbon as shown in (a) at $t/V=0.022$ for $\lambda/J_x=1,2$ and $3$. Both in the SET and SPT phases, the spinon-pair correlation shows exponential decay. (c) Ising correlation $|\langle\sigma^{z}_{0}\sigma^{z}_{r}\rangle|$  is measured along the edge. Inside the SET phase $\lambda/J_x=1,2$, the correlation is short-ranged, meaning symmetric edge without vison condensation. Inside the SPT phase, $\lambda/J_x=3$, the correlation is long-ranged, meaning an Ising symmetry-breaking edge via vison condensation.}
\label{fig:fig6}
\end{figure}

\subsection{SET-SPT transition} 
Next we pay more attention to the topological nature of the SET-SPT phase transition. 

The transition is first-ordered as the energy density has a cusp when we tune $\lambda/J_x$, this is shown in Fig.~\ref{fig:fig3} for the scans of $\lambda/J_x$ with $t/V=0.022$ and $0.02$, respectively. 

The phase boundary between the SET and the SPT phases can be also determined by the bulk Ising-odd vison-pair correlator $\langle v^{o}_{I}v^{o}_{J}\rangle$ in Eq.~\eqref{eq:visoncorrelator}. The path $C$ of the correlator is shown in Fig.~\ref{fig:fig4} (a). This correlator is short-ranged in the SET phase and long-range ordered in the SPT phase. The corresponding results are shown in Fig.~\ref{fig:fig4} (c) at $t/V=0.022$ for various $\lambda/J_x$ across the SET-SPT transition. On the other hand, there is no long-range order in the bulk Ising-Ising correlator $|\langle \sigma^z_0 \sigma^z_r \rangle|$, as shown in Fig.~\ref{fig:fig4} (b), both in the SET and SPT phase. This again means that the bulk is gapped in the Ising channel of both SET and SPT phases.
 
As discussed in Sec.~\ref{sec:measurable}, the SET-SPT transition can be captured by a topological index. To this end, we first introduce a pair of Ising defects separated by distance $l=L/2$ into the system and measure the topological index $Z$ in Eq.~\eqref{eq:bosonnumber} as shown in Fig.~\ref{fig:fig5}(a) in a region $\mathcal{D}$ circulated by the black solid line in Fig.~\ref{fig:fig5} (a) around one Ising defect. As the Ising defect in SPT phase carries fractional charge, when we fix $t/V=0.022,J_x=0.05$ and increase $\lambda/J_x$ to trigger the SET to SPT transition, a jump in $Z$ is expected across the transition point. As mentioned before, in our QMC simulations, $V/t$ is not infinite, and we have projected the wave-function onto the low-energy subspace where 3-boson per plaquette constraints are exactly satisfied before taking the expectation value. In Fig.~\ref{fig:fig5} (c), we can see that $Z$ jump from $-1$ to $+1$ when we increase $\lambda/J_x$ with $t/V=0.022$ and $J_x=0.05$. This is consistent with SET-SPT phase transition $\lambda/J_x\sim 2.0$ obtained from the bulk Ising-odd vison-pair correlator $\langle v^{o}_{I}v^{o}_{J} \rangle$ shown in Fig.~\ref{fig:fig5} (b). The three non-integer points, between $Z=-1$ and $Z=1$, are due to finite size effect. 

\subsection{Transition and Ising symmetry breaking on the boundary} 
The topological index represents the bulk signature of the SET-SPT transition, there are signatures from the edge as well. As discussed in Sec.~\ref{sec:measurable}, across the SET-SPT transition, one expects along the edge there will be an Ising transition in the vison-vison correlator, namely, the correlation is short-ranged in the SET phase, but develops long-range order in the SPT phase. To observe such behavior, we performed simulation of ribbon geometry with periodic boundary only along the $\mathbf{r}_{1}$ direction of the kagome lattice, as shown in Fig.~\ref{fig:fig6} (a), and the ribbon is $L_{x}=12$ long and $L_y=12$ wide. 

We first measure the spinon-pair correlation along the edge, it is given by $\langle b^{\dagger}_{0}b_{r} \rangle$, as shown in our previous work~\cite{GYSun2018}, spinon-pair is gapped in the SET phase, and since the SET-SPT is achieved by condensing the visons instead of spinons, in the SPT phase the spinon excitations are gapped as well. Because of this, in both sides of the transition, spinon-pair correlation along the edge shows exponential decay. This is the case of Fig.~\ref{fig:fig6} (b), where for $\lambda/J_x=1,2$ and $3$ (note the first order bulk transition is around $\lambda/J_x \sim 2.0$) in a semi-log plot, the correlations show straight lines, signifying the exponential decay of spinon-pair correlation along the edge in both phases.

Then we measure the local Ising-Ising correlator $|\langle\sigma^{z}_0\sigma^{z}_r\rangle|$ along the edge, as shown in Fig.~\ref{fig:fig6} (c). For $\lambda/J_x \le 2$, this correlator is short-ranged, reflecting the fact that there is no vison condensation in the SET phase. But for $\lambda/J_x=3$, once the bulk is inside the SPT phase, this correlation develops long-range order. This shows that across the SET-SPT transition, because the condensation of the Ising-odd visons, there is a Ising symmetry breaking along the edge. It is also interesting to note that although the edge develops Ising long-range order, but the Ising-Ising correlator $|\langle\sigma^{z}_0\sigma^{z}_r\rangle|$ is short-ranged in the bulk, as shown in Fig.~\ref{fig:fig5} (b), consistent with the topological nature of SPT phase.

\subsection{A mean-field theory for the phase diagram} \label{sec:mean-field}

In the phase diagram Fig.\ref{fig:fig1}(b) we observed that the second-order SF-SET phase transition line is terminated at the first order SF-SPT and SET-SPT transitions, leaving a critical endpoint where all three phases meet. It is interesting to understand whether these are generic properties of the phase diagram. In fact, based on a simple mean-field theory below, these are indeed universal behaviors of the phase diagram adjacent to the SF-SET second-order line. 

\begin{figure}[htp!]
\centering
\includegraphics[width=\columnwidth]{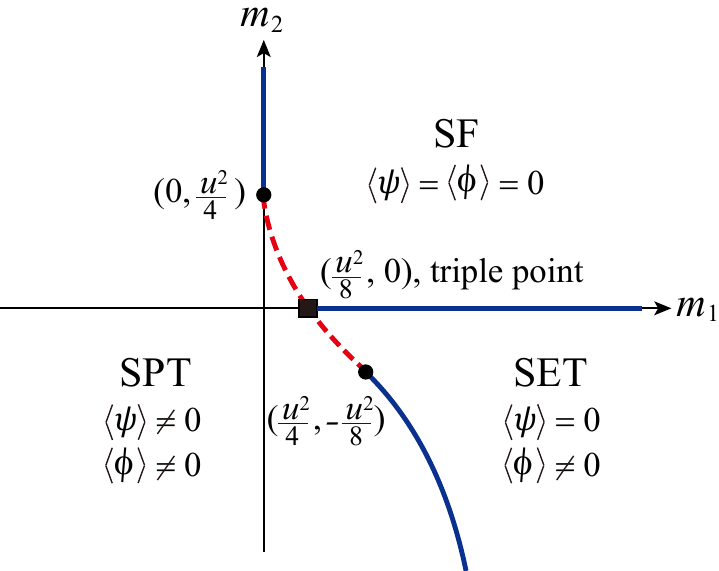}
\caption{The mean-field phase diagram of Eq.~\eqref{eq:mean-field}. Blue lines: second-order phase transition. Dashed red lines: first-order phase transitions. Black square: critical endpoint. Black dots: tri-critical points separating second-order and first-order transitions.}
\label{fig:mean-field}
\end{figure}

Based on the well-known boson-vortex duality\cite{1978AnPhy.113..122P,PhysRevLett.47.1556,PhysRevB.39.2756}, the superfluid(SF) to SPT phase transition can be interpreted as the condensation of single vortex, and the SF to SET phase transition can be interpreted as the condensation of double vortex. Note that comparing with a usual SF to Mott insulator phase transition, here the SF-SPT phase transition requires the condensed single-vortex carrying odd $Z_{2I}$ Ising charge. But this does not modify the effective theory for the SF-SPT transition. These facts motivate us to consider the following Lagrangian

\begin{eqnarray}
    L&=&\int d^3x|(\partial-ia)\psi|^2+m_1|\psi|^2+|\psi|^4+|(\partial-2ia)\phi|^2 \nonumber\\
    &+&m_2|\phi|^2 + |\phi|^4 +\frac{u}{2}[\psi^2\phi^{*}+(\psi^{*})^2\phi],
\end{eqnarray}

where $\psi$ field represents the single vortex field, $\phi$ field represents the double vortex field, $a_\mu$ is the emergent $U(1)$ gauge field and we have rescaled $\psi$ and $\phi$ fields such that $|\psi|^4$ and $|\phi|^4$ terms have coefficient 1.

We then use the mean-field approximation by ignoring the fluctuations of $a$, $\phi$ and $\psi$ fields. The phase factors can be chosen such that $\phi$ and $\psi$ are both real fields. To minimize the free energy, $\phi$ and $\psi$ are translation-invariant. In this approximation, the leading-orders free energy density is simply:
\begin{equation}
    f=m_1\psi^2+\psi^4+m_2\phi^2+\phi^4 +u\psi^2\phi.
\label{eq:mean-field}
\end{equation}

The structure of the mean-field phase diagram can be determined by taking several limits. First, when $m_1,m_2\ll0$, we know that both $\psi$ and $\phi$ fields will be condensed. This corresponds to the SPT phase. Next, when $m_1\gg0$ and $m_2\gg0$, both $\psi$ and $\phi$ are un-condensed. This corresponds to the superfluid phase. And when $m_1\gg0 $ and $m_2\ll0$, $\psi$ is un-condensed and $\phi$ is condensed. This corresponds to the SET phase.

Detailed structure of mean-field phase diagram can be analytically obtained by minimizing the free energy and examining the condensation pattern of $\psi$ and $\phi$ fields (see Appendix \ref{app:mean-field} for details), which is plotted in Fig.~\ref{fig:mean-field}. 

Remarkably, this mean-field phase diagram reproduces all the features of observed phase diagram Fig.\ref{fig:fig1}. In particular, as long as the coupling between the single-vortex and double-vortex $u$ is nonzero (which is generically true), the second-order SF-SET line must terminate at the first-order SF-SPT and SET-SPT transition lines via a critical endpoint. This indicates that the observed nature of the phase diagram Fig.\ref{fig:fig1} is rather universal. In addition, this mean-field phase diagram implies that in order to observe a second-order SET-SPT phase transition, one must search for parameter regimes far away from the SF-SET second-order boundary. We leave this interesting possibility as a topic of future research.

\section{Discussions}
In this work, we performed large-scale quantum Monte Carlo simulatoin to unveal the phase diagram of a quantum spin model which is subject to the dyonic Lieb-Shultz-Mattis theorem. 
The theorem predicts there are SET and SPT phases in the phase diagram. Our numerical results find a first order quantum phase transition between SET and SPT phases, consistent with an anyon condensation mechanism that enforces SPT phase according to the dyonic LSM theorem. Also there is symmetry-breaking SF phase in the phase diagram. The transition between SET and SF is continuous and that between SPT and SF is first order. The SET-SPT transition acquires several non-trivial measurable consequencies such as the bulk vison-pair correlation, the topological index as well as the edge Ising symmetry-breaking at the transition, all these features are captured by our QMC results. Moreover, a mean-field analysis of vortex fields coupled to emergent $U(1)$ gauge field explains that the structure of our obtained phase diagram is universal.

Our work is not only the first unbiased verfication of the dyonic LSM, more importantly, our approach, combinding theoretical understanding such as dyonic LSM and large-scale numerical simulation such as QMC, demonstrate a systematical as well as practical way to generate bosonic SPT phases. We believe that such an approach and the results obtained here pave the way of controlled investigations of phase transitions in the quantum matter beyond Landau-Ginzburg-Wilson paradigm and the novel phenomena, such as anyon condensations leading to SET-SPT phase transitions.

\section*{Acknowledgement}
The authors thank Yang Qi, Ashvin Vishwanath, Chenjie Wang for helpful discussions. YCW acknowledges fundings from the  National Natural Science Foundation of China under Grant No. 11804383, from the Natural Science Foundation of Jiangsu Province under Grant No. BK20180637, and from Fundamental Research Funds for the Central Universities under Grant No. 2018QNA39. ZYM acknowledges fundings from the Ministry of Science and Technology of China through National Key Research and Development Program under Grant No. 2016YFA0300502, from the Strategic Priority Research Program of Chinese Academy of Sciences under Grant No. XDB28000000 and from the National Science Foundation of China under Grant Nos. 11421092, 11574359 and 11674370. XY and YR acknowledge support from the National Science Foundation under Grant No. DMR-1712128. We thank the Center for Quantum Simulation Sciences at Institute of Physics, Chinese Academy of Sciences, the Tianhe-1A platform at the National Supercomputer Center in Tianjin for technical support and generous allocation of CPU time.
\bibliography{bspt}

\appendix
\section{{Perturbation theory study of $H_{\text{eff}}$}}
\label{app:Heff}
In this section we derive the low-energy effect Hamiltonian of the decorated BFG model in the parameter regime where $V,\lambda \gg t,J$.

We separate the Hamiltonian in two parts $H=H_0+H_p$, with
\begin{equation}
H_0=\sum\limits_{( ij)}V_{ij}n_in_j-\sum\limits_{i}\lambda(n_i-1/2)s_{IJ}\sigma^z_I\sigma^z_J.
\end{equation}
and
\begin{equation}
H_p=\sum\limits_{( ij)}(-t_{ij}b^{\dagger}_ib_j+h.c.)-\sum\limits_{\langle IJ\rangle}J_x \sigma^x_I\sigma^x_J.
\end{equation}

Here the eigen-states of $H_0$ are known and the ground state is highly degenerate. And the typical energy scale associated with $H_p$ is much smaller than the energy gap of $H_0$ since $t,J_x\ll V,\lambda$. Therefore we can treat $H_p$ as a perturbation to $H_0$ which lifts the ground state degeneracy.

We can perform the Brillouin-Wigner degenerate perturbation theory to obtain the low-energy effective Hamiltonian. This is done as follows. Suppose the ground state energy of $H_0$ is $E_0$. And we define the projector onto the ground state manifold of $H_0$ as $P_g$, then we have 
\begin{equation}
H_{\text{eff}}=E_0+P_g(H_p+H_pG_0'H_p+H_pG_0'H_pG_0'H_p+\cdots)P_g,
\end{equation}
where $G_0'=P_e(E_0-H_0)^{-1}P_e$ and $P_e$ is the projector onto the excited states of $H_0$ and $H_{eff}$ is determined by the lowest order non-constant term on the RHS of the above equation.

In the following, we denote $N$ as the number of Kagome puaquettes.
%Violation of an gluing bond will increase the energy by \lambda

We now calculate the effective Hamiltonian order by order:
\begin{enumerate}
	\item $H_{\text{eff}}^{(1)}=P_gH_pP_g=0.$
	\item $H_{\text{eff}}^{(2)}=P_gH_pG_0'H_pP_g=-N\cdot \frac{9t^2}{V+2\lambda}-3N\cdot \frac{J^2}{4\lambda}$. The first term comes from the process of a boson hopping to an empty site and then coming back. The second term comes from the process of fliping a pair of Ising DOF twice. To this order, we only have constant terms.
	\item $H_{\text{eff}}^{(3)}=P_gH_pG_0'H_pG_0'H_pP_g=-4t^2J(2\cdot\frac{1}{(V+2\lambda)\cdot 4\lambda}+\frac{1}{(V+2\lambda)^2})\cdot H_{eff}^t-\frac{t^3}{(V+2\lambda)^2}\cdot (H_{eff}^v+36N) $. The $H_{eff}^t$ and $H_{eff}^v$ terms are non-constant terms. The $H_{eff}^t$ term corresponds to the bow-tie hopping term in the orginal BFG model and is represented as
	\begin{equation}
	H_{eff}^t=\sum_{\bowtie}(\big|\raisebox{-4mm}\pgelB\big\rangle \big\langle\raisebox{-4mm}\pgelA\big|+h.c.),
	\end{equation}
	where blue/empty circle means 1/0 boson and the red circle means there can be either 1 or 0 boson.
	
	The $H_{eff}^v$ term is a potential term measuring the number of bosons in a triangle which is just the 3rd order term in the original BFG model. And the constant term $-36N t^3/(V+2\lambda)^2$ comes from 1-boson hopping or 2-boson exchange within the same plaquette.
	
	In order to have an exact mapping between the BFG model and the decorated-BFG model, the $H_{eff}^v$ term should be ignored, therefore we further require that
	\begin{equation}
	\begin{split} 
	&t^2J\cdot(\frac{2}{(V+2\lambda)\cdot \lambda}+\frac{4}{(V+2\lambda)^2})\gg \frac{t^3}{(V+2\lambda)^2}\\&\Leftrightarrow J(8+2\frac{V}{\lambda})\gg t.
	\end{split}
	\end{equation}
\end{enumerate}

In summary, the low-energy effective Hamiltonian of the decorated BFG model in the regime where $J_x(8+2\frac{V}{\lambda})\gg t $ and $V,\lambda \gg J_x,t$ is
\begin{equation} H_{\text{eff}}=-J_{\text{ring}}\sum_{\bowtie}(\big|\raisebox{-4mm}\pgelB\big\rangle \big\langle\raisebox{-4mm}\pgelA\big|+h.c.)+\text{const.},
\end{equation}
where
\begin{equation}
J_{\text{ring}}=\frac{2t^2J(V+4\lambda)}{(V+2\lambda)^2\lambda},
\end{equation}

\section{Solving the effective Hamiltonian $H_{\text{eff}}$}\label{app:solveHeff}
In this section, we show that $H_{\text{eff}}$ has a unique symmetric gapped ground state on torus, demonstrating that the ground state is a symmetric short-range-entangled state. 

Our method is to solve this Hamiltonian via a mapping between the low energy Hilbert space of $H_{\text{eff}}$ and that of $H^{\text{BFG}}_{\text{eff}}$.

We define the following map
\begin{equation}
\mathcal{P}: (\ket{\{S_i^z,+\}}+\ket{\{S_i^z,-\}})/\sqrt{2}\rightarrow\ket{\{S_i^z\}},
\end{equation}
which is an isometry between the Ising-even low energy sector of $H_{\text{eff}}$ and the low energy Hilbert space of $H^{\text{BFG}}_{\text{eff}}$.

It can be easily verified that
\begin{equation}
\mathcal{P}H_{\text{eff}}\mathcal{P}^{-1}=H^{\text{BFG}}_{\text{eff}},
\end{equation}
with $J_{ring}$ set equal on the two sides.

Note that the isometry $\mathcal{P}$ can be viewed as a unitary mapping from the Ising-even low energy sector of $H_{\text{eff}}$ onto a specific topological sector of $H^{\text{BFG}}_{\text{eff}}$ on a torus: due to the constraint $2S_i^z(s_{IJ}\sigma^z_I\sigma^z_J)=1$, $\prod (2S_i^z)$ around any loop is fixed by $s_{IJ}$. Because $H^{\text{BFG}}_{\text{eff}}$ is completely gapped inside a specific topological sector, $H_{\text{eff}}$ is also gapped in the Ising-even sector. The ground state $\ket{\psi}$ in the Ising-even sector of $H_{\text{eff}}$, should also be mapped to a ground state $\ket{\psi^{\text{BFG}}}$ of $H^{\text{BFG}}_{\text{eff}}$. 

Next we prove that $\ket{\psi}$ is in fact the ground state in the whole low-energy Hilbert space of $H_{\text{eff}}$. Based on the well-known duality between the Ising model and the $Z_2$ gauge theory\cite{Kogut1979}, the spectrum of the Ising-odd sector of $H_{\text{eff}}$ is above the energy of $\ket{\psi}$ by a finite energy gap. This can be shown as follows. The spectrum in the Ising-odd sector of $H_{\text{eff}}$ can be mapped via a unitary transformation $U=\sigma_I^z$ (where the site $I$ can be arbitrarily chosen) to the Ising-even sector of the modified Hamiltonian $H_{\text{mod}}=UH_{\text{eff}}U^{-1}$. Then under the isometry $\mathcal{P}$, we have
\begin{equation}
\mathcal{P}H_{\text{mod}}\mathcal{P}^{-1}=H^{\text{BFG}}_{\text{mod}}.
\end{equation}

Simple algebra shows that terms in $H^{\text{BFG}}_{\text{mod}}$ is the same as that of $H^{\text{BFG}}_{\text{eff}}$ except that the ring-exchange terms of the three bow-ties enclosing site $I$ are changed in sign. As discussed in Ref.\onlinecite{sheng2005numerical}, the ground state of this Hamiltonian is just a single vison state, which is apparent from the Ising-gauge duality since $\sigma^z_I$ in the Ising model side is identified as vison creation operator in the gauge theory side.

It is numerically verified that the ground state energy of $H^{\text{BFG}}_{\text{mod}}$ is larger than the ground state energy of $H^{\text{BFG}}_{\text{eff}}$ by a finite amount which is identified as the vison energy gap\cite{sheng2005numerical}. Since the unitary transformation $U$ and the isometry $\mathcal{P}$ both preserve the energy spectrum, we know immediately that the Ising-odd sector of $H_{\text{eff}}$ has an energy gap from the ground state $\ket{\psi}$, whose size is identical to the vison energy gap in $H^{\text{BFG}}_{\text{eff}}$. Therefore the ground state of $H_{\text{eff}}$ is $\ket{\psi}$, which is a symmetric SRE state.

\section{Details of the mean-field analysis}\label{app:mean-field}
In this section we do the mean-field analysis to the effective free energy density Eq.\eqref{eq:mean-field}.

The global minima of the free energy density should satisfy $\frac{\delta f}{\delta \psi}=\frac{\delta f}{\delta \phi}=0$. The variation of $f$ with respect to $\psi$ field gives 
\begin{equation}\label{eq:psi}
\frac{\delta f}{\delta \psi}=2m_1\psi+4\psi^3+2u\psi \phi.
\end{equation}

And the variation of $f$ with respect to $\phi$ field gives
\begin{equation}\label{eq:phi}
\frac{\delta f}{\delta \phi}=2m_2\phi+4\phi^3+u\psi^2.
\end{equation}

Solving these two equations will give us the global mean-field phase diagram.

First, we have a second order phase transition line between SPT phase and SF phase located at $m_2\gg 0$ and $m_1=0$. This can be seen as follows. In the parameter regime with $m_2\gg 0,m_1>0$, Eq.\eqref{eq:psi} and Eq.\eqref{eq:phi} has only one solution$(\psi,\phi)=(0,0)$. Therefore the global minimum of free energy density is $f=0$ and this is the SF phase where $\psi$ and $\phi$ are both un-condensed. 

In the parameter regime with $m_2\gg 0,m_1<0$, Eq.\eqref{eq:psi} and Eq.\eqref{eq:phi} has only two solutions, one is $(\psi,\phi)=(0,0)$ and the other is $(\psi,\phi)=(\pm \sqrt{-\frac{u\phi_0+m_1}{2}},\phi_0)$, where $\phi_0$ is the only solution to the equation $4\phi^3+2m_2\phi-\frac{u^2\phi+um_1}{2}$. The global minima of the free energy density is achieved at $(\psi,\phi)=(\pm \sqrt{-\frac{u\phi_0+m_1}{2}},\phi_0)$ with $f= -\frac{m_1^2}{4}+\mathcal{O}(1/m_2)$. In this case $\psi$ and $\phi$ are both condensed, hence it is the SPT phase. It can be readily seen that if we fix $m_2$ and tune $m_1$, we will obtain a second-order phase transition between SPT and SF phase when $m_1$ changes sign since the order parameters change continuously with respect to $m_1$.

When we decrease $m_2$, this second-order phase transition line will join a first-order phase transition line at a tricritical point with $m_1=0,m_2=u^2/4$. This can be seen as follows. The phase transition is essentially triggered by the condensation of $\psi$. Therefore we can rewrite the free energy density in terms of $\psi$ using the relation $\phi=-2\psi^2/u$ on the $m_1=0$ line:
\begin{equation}
f=(-1+\frac{4m_2}{u^2})\psi^4+16\frac{\psi^8}{u^4}+\mathcal{O}(\psi^{10}).
\end{equation}

The tricritical point is therefore located at $m_2=\frac{u^2}{4}$ at which the coefficient of $\psi^4$ become zero.

And we have a second order phase transition line between SPT phase and SET phase located at $m_2\ll 0$ and $m_1=\sqrt{-\frac{m_2u^2}{2}}$. To see this, we study the parameter set $(m_1,m_2)=(\sqrt{-\frac{m_2u^2}{2}}+\delta,m_2)$ with $\delta$ small. For $\delta<0$, we have three local extrema, which are located at $(\psi,\phi)=(0,0)$, $(\psi,\phi)=(0,\pm \sqrt{-\frac{m_2}{2}})$ and $(\psi,\phi)=((\frac{u^2}{16m_2}-\frac{1}{2})\delta,-\sqrt{-\frac{m_2}{2}}- \frac{u}{8m_2}\delta)$ (correct to $\mathcal{O}(\delta^2)$). And the global minimum is achieved at $(\psi,\phi)=((\frac{u^2}{16m_2}-\frac{1}{2})\delta,-\sqrt{-\frac{m_2}{2}}-\frac{u}{8m_2}\delta)$ with $f=-\frac{m_2^2}{4}-\sqrt{-\frac{m_2}{8}}u\delta^2+\mathcal{O}(\delta^3)$. This is the SPT phase with both $\psi$ and $\phi$ condensed.

And for $\delta>0$, we only have two local extrema located at $(\psi,\phi)=(0,0)$ and $(\psi,\phi)=(0,\pm \sqrt{-\frac{m_2}{2}})$. The global minimum of free energy is achieved at $(\psi,\phi)=(0,\pm \sqrt{-\frac{m_2}{2}})$ with $f=-\frac{m_2^2}{4}$. This is the SET phase with only $\phi$ condensed. Therefore we conclude that the phase transition triggered by tuning $m_1$ with $m_2$ fixed is again second-order since the order parameter changes continuously during the phase transition.

This second-order phase transition line also terminates at a tricritical point $(m_1,m_2)=(\frac{u^2}{4},-\frac{u^2}{8})$, which becomes a first-order line when $m_1<\frac{u^2}{4}$. This can also be seen by rewriting free energy density in terms of $\psi$ using the relation $\phi=(-2\psi^2-m_1)/u$ on the $m_2=-2\frac{m_1^2}{u^2}$ line:
\begin{equation}
f=-\frac{m_1^4}{u^4}+(\frac{16m_1^2}{u^4}-1)\psi^4+\frac{32m_1}{u^4}\psi^6+\mathcal{O}(\psi^8).
\end{equation}

The tricritical point can be identified as $m_1=\frac{u^2}{4},m_2=-\frac{u^2}{8}$ at which the coefficient of $\psi^4$ becomes zero. 

Therefore there should be a first-order phase transition line connecting $(m_1,m_2)=(0,\frac{u^2}{4})$ and $(m_1,m_2)=(\frac{u^2}{4},-\frac{u^2}{8})$.

And there's a second order phase transition line between the SF phase and the SET phase at $m_1\gg 0$ and $m_2=0$. In this case $\psi$ is un-condensed and the phase transition is triggered by the condensation of $\phi$ alone, which is apparently second-order. This second order line terminates 
at the previously mentioned first-order line connecting $(m_1,m_2)=(0,\frac{u^2}{4})$ and $(m_1,m_2)=(\frac{u^2}{4},-\frac{u^2}{8})$. The precise location of this critical end-point is $(m_1,m_2)=(\frac{u^2}{8},0)$.

To study this critical end-point in more detail we can perturb this point by slightly deforming $m_1$ and $m_2$ and examine adjacent phases and the nature of the phase transitions between them.

We choose $(m_1,m_2)=(u^2/8+\alpha,\beta)$ where $\alpha$ and $\beta$ are small and comparable to each other, $\alpha\sim \beta$. For $\beta>0$, we can solve $\frac{\delta f}{\delta \psi}=\frac{\delta f}{\delta \phi}=0$ to get two sets of local minima, one is located at $(\psi,\phi)=(0,0)$ (dubbed $S_1$), the other is located at $(\psi,\phi)=(\pm(u/4-\frac{3\alpha+2\beta}{u}),-u/4+\frac{2\alpha+2\beta}{u})$ (dubbed $S_2$), which is accurate to $O(\alpha^2)$. At $S_1$ the free energy density is zero. And at $S_2$ the free energy density is $f=\frac{u^2}{16}(\alpha+\beta)+O(\alpha^2)$. Therefore when $\alpha+\beta<0$, the global minimum is at $S_2$, signaling the condensation of both $\psi$ and $\phi$ fields. And when $\alpha+\beta>0$, the global minimum is achieved at $S_1$, meaning that $\psi$ and $\phi$ fields are both un-condensed. This is a first-order phase transition since the order parameters $\langle \psi\rangle$ and $\langle \phi\rangle$ have a jump across the phase transition. 

For $\beta<0$, the free energy has two sets of local minima located at $(\psi,\phi)=(0,\pm \sqrt{-\beta/2})$ (dubbed $S_1$) and $(\psi,\phi)=(\pm(u/4-\frac{3\alpha+2\beta}{u}),-u/4+\frac{2\alpha+2\beta}{u})$ (dubbed $S_2$). The free energy density at $S_1$ is $-\beta^2/4$ and the free energy density at $S_2$ is again $f=\frac{u^2}{16}(\alpha+\beta)+O(\alpha^2)$. Similar to the previous discussion, when $\alpha+\beta<0$, we have a condensation of $\psi$ and $\phi$ fields, which indicates the SPT phase. And when $\alpha+\beta>0$, we have a condensation of $\phi$ fields, which drives the system into a SET phase. This is apparently a first-order phase transition.

And the phase transition between the superfluid phase and the SET phase is second order as the order parameter $\langle \phi\rangle$ changes continuously across the phase transition when we tune the mass parameter $m_2=\beta$.

The above analysis shows that the SET-SF phase transition is always second order, but the SPT-SET and SPT-SF phase transition are first-order when close to the triple point, which is a robust feature as long as $u\neq 0$. And well away from the triple point the SPT-SF transition and the SPT-SET transition could become second-order.

\end{document}